# Investigating the IBEX Ribbon Structure a Solar Cycle Apart

M. A. Dayeh[1,2], E. J. Zirnstein[3], P. Swaczyna[3], and D. J. McComas[3]

[1] *Southwest Research Institute, San Antonio, TX 78238 (maldayeh@swri.org)*

[2] *University of Texas at San Antonio, San Antonio, TX 78249*

[3] *Department of Astrophysical Sciences, Princeton University, Princeton, NJ, 08544*

**Abstract.**

A 'Ribbon' of enhanced energetic neutral atom (ENA) emissions was discovered by the Interstellar Boundary Explorer (IBEX) in 2009, redefining our understanding of the heliosphere boundaries and the physical processes occurring at the interstellar interface. The Ribbon signal is intertwined with that of a globally distributed flux (GDF) that spans the entire sky. To a certain extent, Ribbon separation methods enabled examining its evolution independent of the underlying GDF. Observations over a full solar cycle revealed the Ribbon's evolving nature, with intensity variations closely tracking those of the solar wind (SW) structure after a few years delay accounting for the SW-ENA recycling process. In this work, we examine the Ribbon structure, namely, its ENA fluxes, angular extent, width, and circularity properties for two years, 2009 and 2019, representative of the declining phases of two adjacent solar cycles. We find that, (i) the Ribbon ENA fluxes have recovered in the nose direction and south of it down to ~25° (for energies below 1.7 keV) and not at mid and high ecliptic latitudes; (ii) The Ribbon width exhibits significant variability as a function of azimuthal angle; (iii) Circularity analysis suggests that the 2019 Ribbon exhibits a statistically consistent radius with that in 2009. The Ribbon's partial recovery is aligned with the consensus of a heliosphere with its closest point being southward of the nose region. The large variability of the Ribbon width as a function of Azimuth in 2019 compared to 2009 is likely indicative of small-scale processes within the Ribbon.

# 1. Introduction

The Interstellar Boundary Explorer (IBEX) mission is dedicated to imaging energetic neutral atom (ENA) emissions generated around the heliospheric interface with the very local interstellar medium (McComas et al., 2009a). Launched in 2008 and equipped with two ENA sensors, IBEX-Hi (Funsten et al. 2009) and IBEX-Lo (Fuselier et al. 2009), respectively covering overlapping energies 0.5-6 keV and 0.01-2 keV, IBEX continues to provide ENA measurements after a full solar cycle worth of observations (McComas et al. 2020). One of IBEX's main discoveries was the "Ribbon" - a band of enhanced ENA emissions that are ~2-3 times larger than the underlying globally distributed flux (GDF), and spans almost across the entire sky (McComas et al., 2009b). The Ribbon appears at energies between ~0.4–6 keV and is most prominent between ~1 and 3 keV. It is a narrow structure with its center lying in the vicinity of the pristine magnetic field direction of the local interstellar medium (LISM; Frisch et al., 2012; Frisch & Schwadron 2014), whose field lines are draped around the heliosphere. Several lines of evidence from IBEX observations and numerical simulations showed that the most likely mechanism for creating the Ribbon is the secondary ENA mechanism (e.g., Heerikhuisen et al., 2010; Gamayunov et al. 2010; Schwadron & McComas, 2013; Chalov et al., 2010; Zirnstein et al., 2013, 2015, 2017; Dayeh et al. 2019). In this mechanism, 'primary' ENAs generated by solar wind ions neutralized in the inner heliosphere travel into the outer heliosheath and undergo two sequential charge exchange events: an ionization event followed by another neutralization event, creating "secondary" ENAs. The time between these charge exchange events is typically described by the 1/e mean free time for charge exchange to occur, which is on the order of ~1 yr ENAs measured by IBEX-Hi. After re-neutralization, some secondary ENAs travel back to 1 au where they are measured by IBEX. Because they are most intense along the line of sight that is perpendicular to the interstellar magnetic field (ISMF) (i.e., $B.R=0$), their measurement partially depends on their original direction of motion, the location of the charge exchange (e.g., Heerikhuisen et al., 2010; Schwadron & McComas, 2013; Zirnstein et al., 2015, 2016b), the interstellar turbulence power spectrum (Burlaga et al. 2018; Giacalone & Jokipii 2015; Gamayunov et al. 2019; Zirnstein et al. 2020), and scattering of the progenitor ions (e.g., Schwadron & McComas 2013; Heerikhuisen et al. 2014; Isenberg 2014, Gamayunov et al. 2017, Zirnstein et al. 2019, 2023).

The Ribbon and GDF signals are superposed from the two different populations. They are both measured by IBEX as an integrated accumulation along IBEX-Lo and -Hi's line of sight (i.e., boresight). Several studies have attempted to separate these populations using different data and modeling assumptions (Schwadron et al. 2011, 2014, 2018; Dayeh et al. 2019, 2023, Swaczyna et al. 2022, 2023; Reisenfeld et al. 2021). However, this task has proven to be non-trivial and strongly depends on the underlying data and modeling (e.g., Ribbon transverse profile) assumptions. The overlapping emission structures of the source thus introduce ambiguity in inferring information about their properties and basic assumptions about ENA sources are unavoidable. Furthermore, low statistics in some directions of the sky and the IBEX spatial and temporal resolutions constrain the separation process. The upcoming Interstellar Mapping and Acceleration Probe (IMAP; McComas et al. 2018) is expected to provide improved ENA temporal and spatial measurements, which will further advance identifying the properties of the Ribbon and the GDF more completely.

The Ribbon and the GDF signals both exhibit distinct temporal variations and respond to solar cycle changes, with the GDF showing a smaller temporal latency than the Ribbon by a few years (McComas et al. 2017, 2020; Schwadron et al. 2018; Swaczyna et al. 2022). These variations have been utilized to distinguish the Ribbon from the underlying GDF, as demonstrated by Dayeh et al. (2023). McComas et al. (2020) discussed and analyzed IBEX ENA imaging of the heliosphere over a solar cycle (11 years), from 2009 to 2019. In particular, they reported that ENA intensities in the Ribbon at 1.1 keV showed signs of recovery in 2019 in directions near the nose of the heliosphere and just below the ecliptic plane (see their Figure 28), after a general dimming occurred earlier in the solar cycle.

The analysis by McComas et al. (2020) motivates this work to investigate the Ribbon structural changes in two adjacent solar cycles (23 and 24), but both in similar phases of their respective solar cycle and with comparable SW dynamic pressure. In this paper, we investigate the properties of the Ribbon in two years separated 10 years apart and lie on the same declining phase of one solar cycle apart. We note that IBEX measurements are made in 2009 and 2019, but due to SW-ENA time delays, the SW conditions effecting these ENA measurements are a few years earlier. Thus the corresponding SW conditions are during the declining phase of solar cycles 23 and 24, respectively. We aim at determining the extent of the Ribbon structural recovery, including its integrated ENA intensity, width, angular distance of the Ribbon peak from the map center, and circularity properties.

## 2. Data, Methodology, and Observations

We use IBEX-Hi ENA data from validated data release #16 (McComas et al. 2020). Data is survival-probability-corrected and is from the ram-only direction. Ribbon-centered maps come from the IBEX Science Operation Center (SOC; Schwadron et al. 2009) and maps at each energy are centered on the Ribbon centers determined from Dayeh et al. (2019), while the nose is located along the 0° azimuth angle. Note that the maps are not obtained through rotation and interpolation of the standard IBEX maps in the ecliptic coordinates, but the ENA events are directly binned into pixels in the rotated frame. We only use years 2009 and 2019 as the goal of this study is to assess the recovery of the Ribbon properties and structure over 11 years of data and not how it evolves during the years. Part of the latter has also been briefly examined by McComas et al. (2020), Zirnstein et al. (2023), and Swaczyna et al. (2023).

Figure 1 shows the ribbon-centered sky maps obtained by IBEX during 2009 and 2019, for five IBEX-Hi energy passbands (columns). Panels (a) and (b) show the full ENA signal of the Ribbon and the GDF and panels (c) and (d) show the Ribbon-only signal. The Ribbon-only maps are obtained as a difference between the observations including both the GDF and Ribbon components and the GDF estimated as a linear combination of low degree spherical harmonics by Swaczyna et al. (2022). Consistent color-coding is shown for each energy passband for easy comparison.

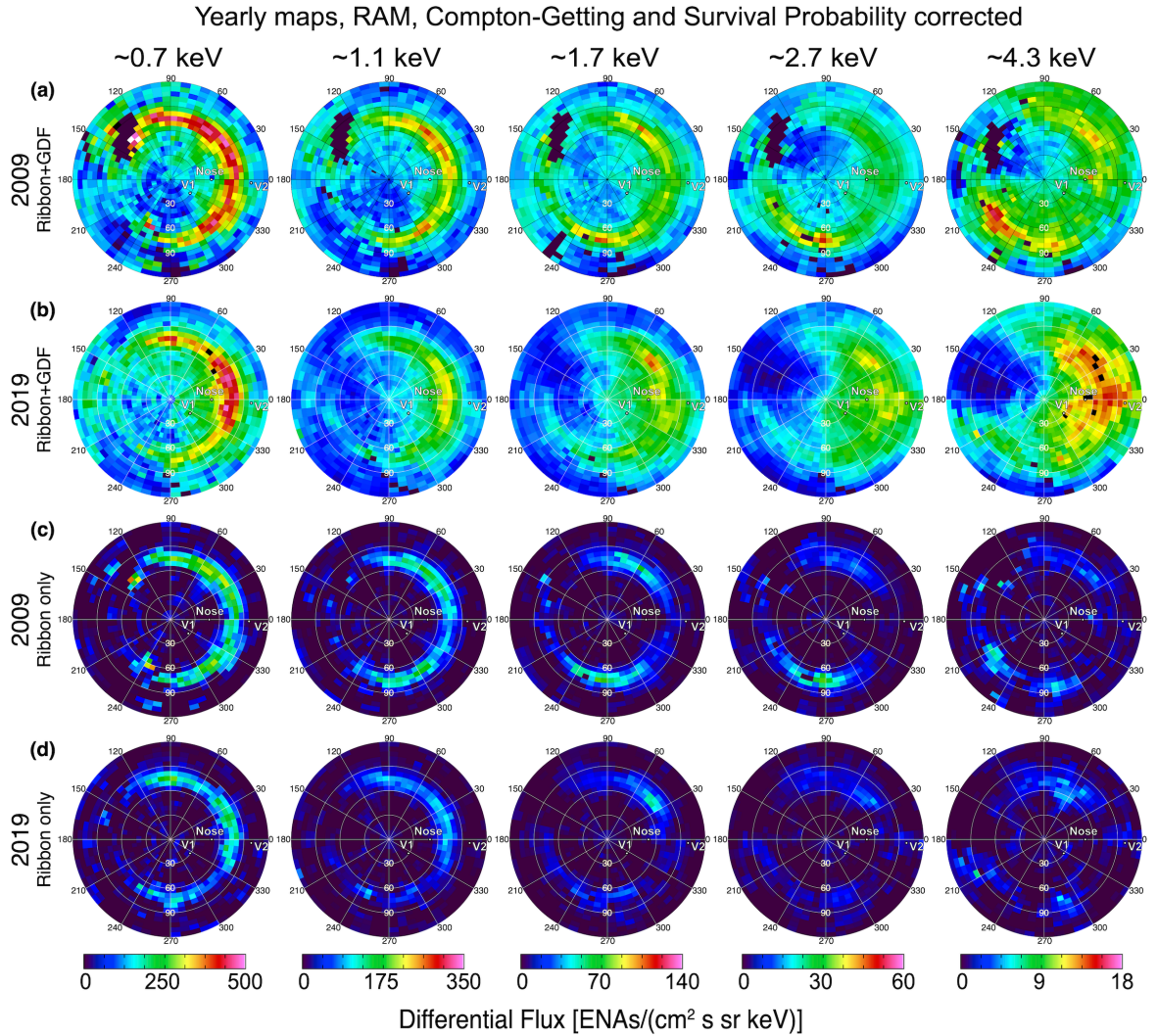

*Figure 1. (a, b) Ribbon-centered maps showing all-sky ENA fluxes at five IBEX-Hi energies for 2009 and 2019. (c,d) Separated Ribbon-only maps for 2009 processed using maps from (a,b) and the Ribbon separation method of Swaczyna et al. (2022). Labels mark the directions of upwind interstellar inflow (nose), and the Voyager spacecraft (V1 and V2).*

Large asymmetric emissions are apparent in both types of maps. Significant time evolution in the combined GDF and Ribbon becomes clear in higher energy steps ~2.7 keV and 4.3 keV, where the position of the maximum flux is in a different portion of the map (closer to the azimuth of ~210-270° in 2009, and ~0-45° in 2019). Nonetheless, it is still hard to isolate the evolution of the flux components. While the enhancement at the azimuthal angle of ~220° in 2009 is visible in both the Ribbon separated and total flux maps, the enhancement near the azimuthal angle of ~0° in 2019 in the total map is not visible in the separated Ribbon. The increase in 2019 is an expanding global feature in response to the increase of the solar wind dynamic pressure (McComas et al. 2019, 2020; Zirnstein et al. 2022). Panels (c) and (d) show that there is an overall drop in the Ribbon-only fluxes, most pronounced at lower energies with an asymmetric behavior. For instance, fluxes at ~1.1 keV and especially ~1.7 keV in 2009 are enhanced in both directions away from the nose, i.e., going Northward (Clockwise) and Southward (anticlockwise). This symmetry below 1.1 keV disappears in 2019, showing asymmetric enhancements appearing only on one-side (mostly southward). Furthermore in 2009 and at ~1.7 keV, the Ribbon is most pronounced northward of the nose. However, this flips in 2019 and the most enhanced region becomes southward of the nose. These asymmetries in the latitudinal structure of the Ribbon flux could be reflections of the asymmetric evolution of the polar coronal holes from in solar cycles 23 and 24 (Sokół et al. 2020, Zirnstein et al. 2020). It could also be that we are not yet seeing the northern ribbon response. Secondary ENA models predict that the ribbon's source is farther from IBEX in the high northern latitudes compared to the mid southern latitudes (e.g., Zirnstein et al. 2019, Figure 4). This is also supported by the response of the GDF to the large SW pressure enhancement in solar cycle 24; the southern ribbon overlaps the section of the sky where the heliosphere boundaries are closest to the Sun, where the responses of the GDF to changes in the SW pressure were first observed (McComas et al. 2018, 2019).

To provide a quantitative comparison of Ribbon-only fluxes, we examine their evolution in each azimuth of the sky. Figure 2a shows fluxes in 2009 (in blue) and 2019 (in red) for a selected swath in the sky (direction at 39°). Fluctuations near zero are statistically non-significant and indicate the ENA Ribbon background level. To compare both years, we first identify Ribbon boundaries where the enhancement above the fluctuating background appears.

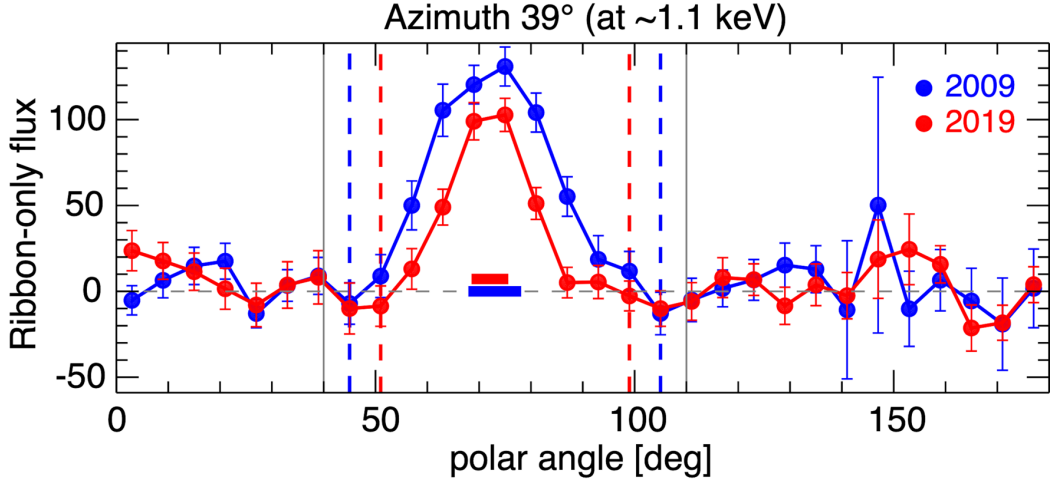

*Figure 2. Illustration of Ribbon region determination used to assess its recovery. Ribbon-only fluxes from 2009 (blue) and 2019 (red). Vertical dashed lines indicate the derived boundaries of the ribbon. Horizontal bars show the Ribbon widths around the center (see text). We determine the Ribbon-only properties and examine their recovery in different sky directions along the Ribbon.*

We start by identifying a broad region between the polar angles of 45° and 110°, which encompasses the Ribbon at all energies, as indicated by the vertical grey lines. For each year, we then locate the pixel with the maximum ENA flux (i.e., Ribbon peak) and find the pixels nearest to the zero-level that are closest to this peak pixel on both sides of the Ribbon. In other words, going down the Ribbon 'hill' on both sides until the zero-threshold level is met. Here, the 'nearest to zero' level is defined when this condition is satisfied,

$$J_{loc} + \frac{\sigma_{J_{loc}}}{2} \leq 0 \tag{1}$$

Where $J_{loc}$ is the local flux of the Ribbon and $\sigma_{J_{loc}}$ is its uncertainty. The term proportional to the uncertainty is included to avoid the situation in which a negative bin is removed due to statistical fluctuation. We use this limited range of polar angles to reduce impact of the flux variations outside of the Ribbon, which would impact the momentum method implemented in this study.

Throughout the analysis, we have also excluded azimuths from the Ribbon-separated data that have originally missing pixels in the Ribbon region. We then use the remaining Ribbon-only regions for all energies and azimuths to assess the Ribbon structure recovery by deriving the three quantities defined below.

*i. Ribbon Integrated Flux; $J_R$:* This quantity measures the total flux under the Ribbon within the determined boundaries, as:

$$J_R = \sum_i J_i \qquad (2)$$

and its propagated uncertainty is:

$$\delta_{J_R} = \sqrt{\sum_i \delta_{J_i}^2} \qquad (3)$$

We opted to use the Ribbon integrated fluxes and not the peak ribbon flux because of the non-uniformity and the apparent asymmetric recovery of the Ribbon. The integrated flux thus better describes the changes of the total ENA intensity within the Ribbon.

*ii. Ribbon Angular Distance from the map center; $\varphi_R$:* This also indicates the Ribbon mean location away from the map center. Here, we define it as the first moment ($I_1$) of the Ribbon profile and is thus weighted by the ENA fluxes within, given by,

$$I_1 = \varphi_R = \frac{\sum_i \theta_i \cdot J_i}{\sum_i J_i} \qquad (4)$$

where $j_i$ is the derived Ribbon flux at each polar angle $\theta_i$.

The corresponding uncertainty is,

$$\delta_{I_1} = \sqrt{\frac{\sum_i \left((\theta_i - \varphi_R)^2 \, \delta_{J_i}^2\right)}{(\sum_i J_i)^2}} \qquad (5)$$

*iii. Ribbon width, $\sigma_R$:* This quantity measures the 1-sigma width of the Ribbon, and is derived using the second central moment ($I_2$) of the Ribbon distribution, such as:

$$I_2 = \frac{\sum_i \theta_i^2 J_i}{\sum_i J_i} \qquad (6)$$

and the Ribbon width, $\sigma_R$, is given by the second central moment,

$$\sigma_R = \sqrt{I_2 - \varphi_R^2} \tag{7}$$

The associated uncertainty, $\delta_{\sigma_R}$, can be written as,

$$\boldsymbol{\delta_{\sigma_R}} = \frac{1}{2 \cdot \sigma_R} \sqrt{\frac{\sum_i \left( \delta_{J_i}^2 (\theta_i^2 - I_2 + 2I_1^2 - 2\theta_i I_1)^2 \right)}{(\sum_i J_i)^2}} \tag{8}$$

In Figure 3, we plot the above derived Ribbon quantities (numbered *i* to *iii* above) at all energies and for both examined years. The figure has three plates (P1, P2, and P3) with 5 panels each showing the IBEX-Hi energy passbands (a though e). Here, we plot all properties in a single Figure so that it is easier to view the changes in all quantities at once. In this Figure, data gaps indicate additional excluded regions such as the heliotail (e.g., Dayeh et al. 2022), the magnetosphere obstacle (e.g., McComas et al. 2014) and regions where the Ribbon could not be determined.

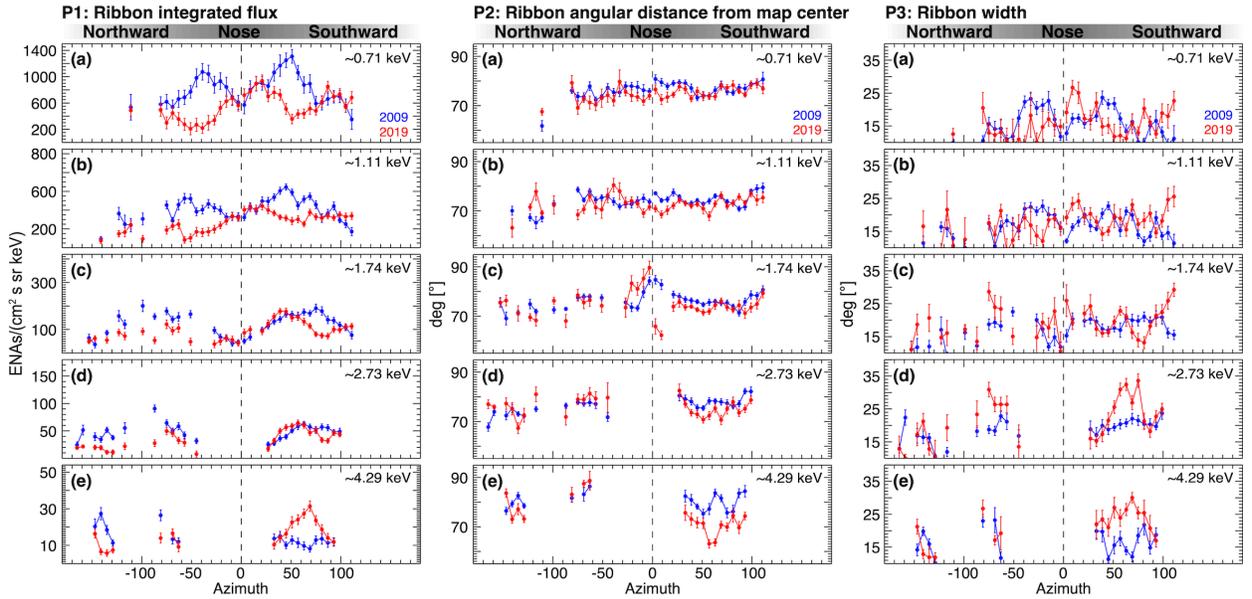

*Figure 3. Derived properties of the Ribbon-only fluxes comparing 2009 and 2019: (P1) Integrated flux, (P2) angular distance from the map center, and (P3) width. Panels from top to bottom show results for observed energy steps. Data gaps indicate regions excluded due to the lack of Ribbon in these areas (see text). Trends show very interesting behavior of the Ribbon's difference. Most importantly that it varies at different rates away from the nose.*

We explain the behavior in Figure 3 as a function of energy as the variability seems to be organized mostly as such. Two main results emerge.

At low energies (~0.7 and ~1.1 keV), the integrated flux $J_R$ recovers near the nose with similar values during 2009 and 2019. This recovery is less pronounced in longitude from the nose toward southern (beyond ~25° azimuth) and northern (beyond ~15° azimuth) latitudes, with the largest difference appearing near ~45° away from the nose on both sides (Figure 3, P1a, b). Beyond ~60°, fluxes from the southward direction become comparable again. Note that in ecliptic coordinates, this translates to a region between ~25°-30° south of the ecliptic plane. The angular distance, $\varphi_R$, however, is consistently a little larger in 2009 compared to 2019 closer to the nose. The width $\sigma_R$ is more variable; both 2009 and 2019 trends appear quasi-periodically every ~90°. It thus appears that there may be a global "out of phase" shift in the width behavior of both maps.

At mid energies (~1.7 keV), $J_R$ behaves similarly to lower energies but $\varphi_R$ in 2019 appears larger in the northward direction (within ~30° of the nose) and smaller in the southward direction. $\sigma_R$ at this energy fluctuates largely as well (see P2c). At angles beyond 90°, the width $\sigma_R$ shows the largest difference between 2009 and 2019 at energies below ~1.7 keV, as trends from both years appear to depart away from each other (P3a,b,c).

The highest energies (~2.7 keV and ~4.3 keV) show significant differences in the $J_R$ behaviour, where integrated fluxes are similar between 2009 and 2019 at ~2.7 keV but largely differ at ~4.3 keV. In fact, the 4.3 keV integrated flux in 2019 is significantly larger than that of 2009 and shows a triangular-like formation, peaking near 70° southward (P1d). The angular distance $\varphi_R$ is larger in 2009 compared to 2019 at both energies, but the width $\sigma_R$ is larger in 2019 compared to that of 2009.

Figure 4 illustrates comparisons of the Ribbon angular distance from center and the Ribbon width from a different perspective. Figure 4a is for 2009 and shows the Ribbon width (blue wedges) and the central peak location of each wedge. Yellow error bars represent the statistical uncertainties of the central peak location. As with any separation method, imperfections may introduce systematic errors that are not accounted for in here. Figure 4b shows the same quantities but for 2019, and 4c shows results from both years superposed on top of each other, where the purple color indicates their overlap. Compared to 2009, the 2019 plots (middle column) show that the Ribbon has recovered in nearly all populated directions but at significantly different recovery rates. At the lowest two energy passbands, the Ribbon appears to be most recovered near the nose. Figure 4d shows that the 2019 Ribbon width exceeds that of 2009 between ~50° and 75°, while it remains narrower around this region. Note that a wider Ribbon does not necessarily mean more recovery; looking at Figure 3 in this context, we find that the integrated fluxes are similar in this region (P1d), thus meaning that the peak fluxes must be lower to account for a wider Ribbon.

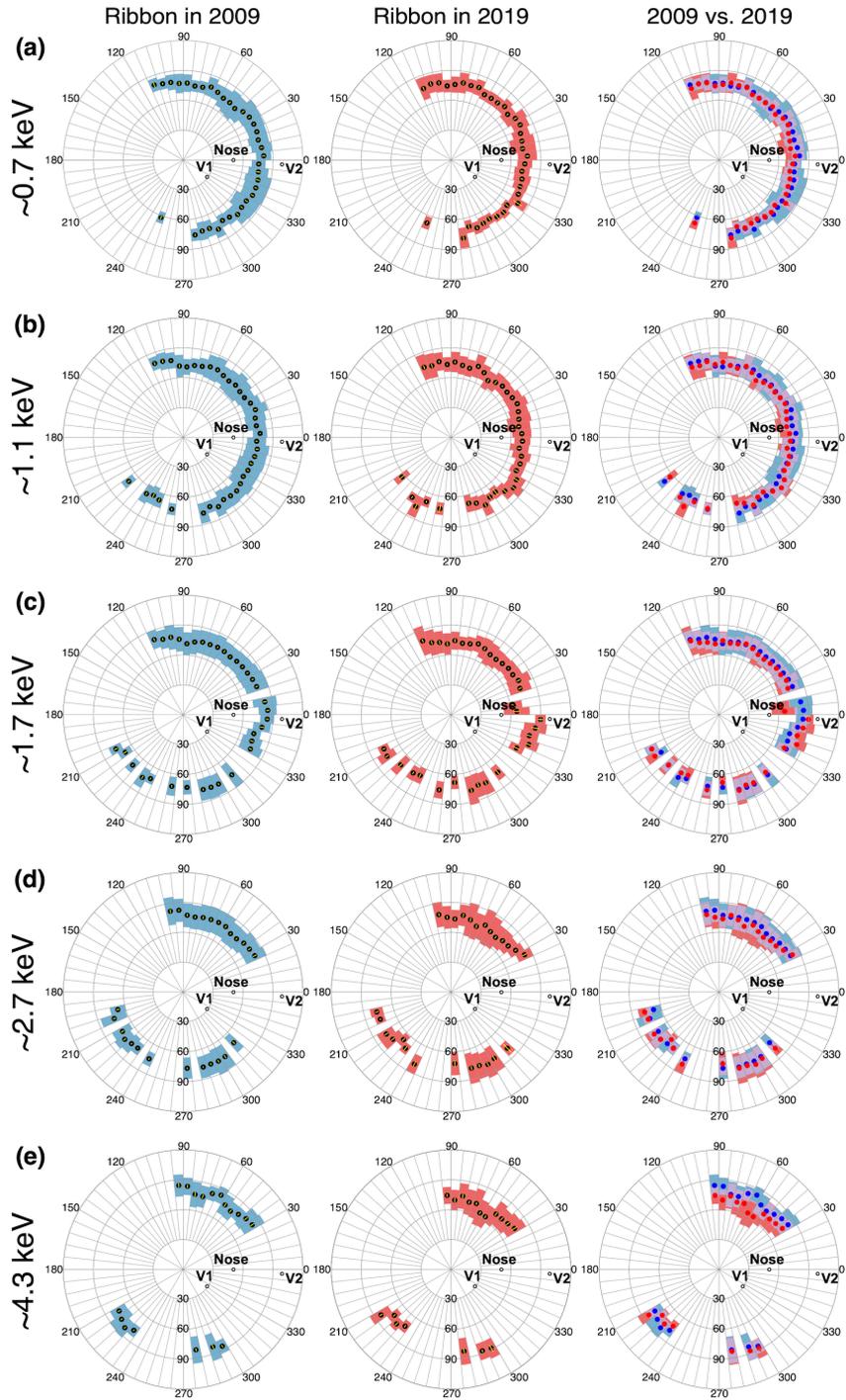

*Figure 4. Ribbon mean and width (i.e., "wedge") locations plotted on a Ribbon-centered map for 2009 (left column), 2019 (middle column), and for both years (right column). In (c), the purple color indicates the overlap between the widths from (a) and (b), easily visualizing where*

*the Ribbon has moved and where it has not. Black dots and superimposed yellow error bars indicate the derived mean angular distances from the center and its uncertainty.*

Our next step involves conducting a circularity analysis on the Ribbon for both periods. To carry out this analysis, we utilize chi-square minimization to fit a circle to the Ribbon peak locations and their uncertainties, as described in Swaczyna et al. (2016), and Zirnstein et al. (2023). The chi-square minimization has the form:

$$\chi^2(\Omega, r) = \sum_{i=1}^{N} \frac{[g(\Omega_i, \Omega) - r]^2}{\sigma_i^2}, \tag{9}$$

with $\Omega = \Omega(\theta_e, \phi_e)$ being the ribbon center in ecliptic J2000 longitude and latitude, $r$ is the derived ribbon radius, and $\sigma_i$ is the uncertainty of the Ribbon peak position. Function $g(\Omega_i, \Omega)$ gives the angular distance of the ribbon along azimuth $i$ from the ribbon center. 1-sigma uncertainties of the derived Ribbon center longitude and latitude are derived from the square root of the diagonals of the covariance matrix after minimization of Equation (9). We also scale these uncertainties by the square root of the reduced chi-square, assuming that the Ribbon can be described by a Gaussian function. The uncertainty of the Ribbon radius is computed by dividing the unweighted standard deviation of the Ribbon peak positions from the mean ribbon radius by $\sqrt{N-1}$, resulting in the unweighted standard error.

Results of the change in the Ribbon circularity parameters between 2009 and 2019 are shown in Figure 5 for four energies as the ~4.3 keV channel did not give a converging fit. Figure 5a shows that the center of the Ribbon shifts to lower longitudes over time for the lowest 3 energy channels, but to a higher longitude for ~2.7 keV. Similarly for latitude, the center in 2019 is at a higher latitude (0.7 and 1.1 keV) and at a lower latitude for 2.7 keV, while the ~1.7 keV Ribbon center latitude does not appear to change between 2009 and 2019. Figure 5b shows that the radius of the Ribbon in 2019 is systematically smaller than that of 2009, although they are still statistically consistent based on our uncertainty analysis. Even though the ~4.3 keV channel is excluded from this figure, changes in Figure 4e point to the same behavior of the Ribbon, where the recovered Ribbon in 2019 is generally closer to the center of the map, compared to that of 2009.

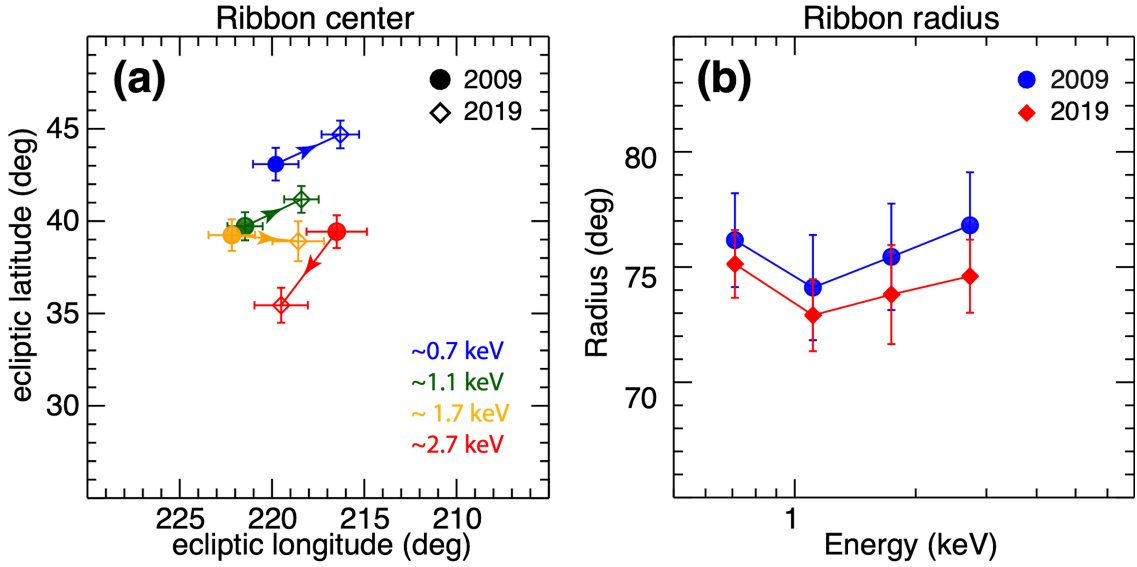

*Figure 5. Results of the Circularity analysis. (a) Variations of the derived center from 2009 to 2019. Arrows indicate the shifting direction. (b) Variations of the derived radius for 2009 (blue) and 2019 (red) at the four ENA energies. Although 2019 radius is systematically smaller than that of 2009, both radii are statistically similar.*

## 3. Summary and Conclusions

The IBEX Ribbon exhibits a gradual and continuous evolution over time. It was observed as a bright structure in 2009 that started dimming in ~2012, reached a minimum after ~2014, and then started to brighten again over the last few years in certain parts of the sky (McComas et al. 2020). Detailed analyses of the Ribbon's time-evolution are possible thanks to novel signal separation techniques that enabled separation for individual IBEX maps (e.g., Swaczyna et al. 2022, 2023). In this work, we focused on assessing the recovery of the ribbon's intensity and the change in its structure between individual maps in 2009 and 2019, over different azimuthal sectors of the Ribbon in a polar centered frame, and five energy passbands covering ~0.5-6 keV from IBEX-Hi. We find the following:

1. Below ~1.7 keV, integrated Ribbon fluxes recover in the region from the nose to ~25° azimuth southward (Figure 3, P1a-c).

2. The Ribbon width exhibits significant variability as a function of azimuthal angle around the map center, with "out-of-phase" variability between 2009 and 2019. (P1e, P2e, P3e).

3. Circularity analysis shows that the 2019 Ribbon exhibits a statistically consistent radius with that in 2009 (Figure 5b).

Recovery of the Ribbon's intensity appears most near the nose and southward of it down to about -25°, latitude, consistent with this part of the ribbon, formed outside the heliopause, being produced closest to the Sun (McComas and Schwadron 2014, McComas et al. 2020, Zirnstein et al. 2022). It also may be related to the asymmetric SW structure observed in the last solar cycle (Zirnstein et al. 2021). Note that the nose slightly lies below the equatorial plane as well, so the recovered region of the Ribbon is completely below the solar equatorial plane. The lowest energy ENAs might be expected to show the most delayed recovery due to their slower propagation speeds; however, these ENAs are thought to be generated at closer distances from the heliopause and thus their recycling distance is closer and they may reflect changes in the SW first. Moreover, because the closest part of the ribbon's source is at low latitudes where the SW speed is, on average, <500 km s$^{-1}$ (McComas et al. 2000), the majority of ENAs produced at these latitudes are <2 keV and significant changes in intensity are expected at these energies only. The large variability exhibited by the Ribbon at ~4.3 keV in 2019 may be due to a large GDF component. Note that the Ribbon at this energy is broad and quickly loses its structure being overwhelmed by the apparent GDF signal. Moreover, there is a lack of available separated Ribbon data at this energy, and thus a reasonable conclusion can not be attained at these energies. The dynamic variability in the Ribbon width as a function of azimuth indicates that different regions of the Ribbon's source evolve at different rates, which is not observed in steady-state Ribbon simulations (e.g., Zirnstein et al. 2013, 2018, 2019), and not clearly visible in time-dependent models with a smoothly varying neutral SW/PUI source (Zirnstein et al. 2023). This observation could stem from different processes, including the evolution of solar wind structure between both solar cycles, be a property of the Ribbon fine structure (e.g., McComas et al. 2009, Fichtner 2014) or related to small scale, time-dependent processes within the Ribbon from, e.g., turbulent interactions (Giacalone & Jokipii 2015; Gamayunov et al. 2019; Zirnstein et al. 2020).

In summary, our analysis shows that the Ribbon structure has not fully recovered in intensity between 2009 and 2019. Moreover, the Ribbon angular distance to the center of the map shows no significant change between 2009 and 2019 at most energies and directions, but the Ribbon width shows significant variability that appears uncorrelated for specific azimuthal sectors along the Ribbon a solar cycle apart, although on average are similar. The circularity analysis indicates a statistically similar radius of the Ribbon for both examined periods. The Ribbon's partial recovery is consistent with the consensus of a heliosphere with its closest point being southward of the nose region. The variability in the width as a function of azimuth is potentially indicative of small-scale variations within the Ribbon source region. These results emphasize the importance of understanding time dependent particle interactions with turbulence in future models of the Ribbon. IMAP measurements with improved angular resolution and ENA statistics could provide more details on these small-scale processes affecting the Ribbon.

**Acknowledgements** This work was supported by NASA's Heliophysics Guest Investigators – Open program under grant no. 80NSSC21K0582. Support at Princeton and SwRI was also provided through the IBEX mission under grant 80NSSC20K0719.

**References**

Burlaga, L. F., Florinski, V., & Ness, N. F. 2018, ApJ, 854, 20

Chalov, S. V., Alexashov, D. B., McComas, D., et al. 2010, ApJ, 716, L99

Dayeh, M. A., Zirnstein, E. J., Desai, M. I., et al. 2019, ApJ, 879, 84

Dayeh, M. A., Zirnstein, E. J., Fuselier, S. A. , et al. 2022, ApJS, *261*(2), 27

Dayeh, M. A., Zirnstein, E. J., and McComas, D., J., 2023, ApJ, 879, 84

Fichtner, H., Scherer, K., Effenberger, F., et al. 2014, Astron Astrophys, 561, A74

Funsten, H. O., Allegrini, F., Bochsler, P., et al. 2009, SSRv, 146, 75

Fuselier, S. A., Allegrini, F., Funsten, H. O., et al. 2009, Sci, 326, 962

Frisch, P. C., Andersson, B.-G., Berdyugin, A., et al. 2012, ApJ, 760, 106

Frisch P. C. and Schwadron N. A. 2014 ASP Conf. Ser. 484, ed Q. Hu and G. P. Zank (San Francisco, CA: ASP) 42

Gamayunov, K. V., Heerikhuisen, J., & Rassoul, H. 2017, ApJ, 845, 63

Gamayunov, K. V., Heerikhuisen, J., & Rassoul, H. K. 2019, ApJL, 876, L21

Gamayunov, K., Zhang, M., & Rassoul, H. 2010, ApJL, 725, 2251

Giacalone, J., & Jokipii, J. R. 2015, ApJL, 812, L9

Heerikhuisen, J., Pogorelov, N. V., Zank, G. P., et al. 2010, ApJL, 708, L126

Heerikhuisen, J., Zirnstein, E. J., Funsten, H. O., Pogorelov, N. V., & Zank, G. P. 2014, ApJ, 784, 73

Isenberg, P. A. 2014, ApJ, 787, 76

Reisenfeld, D. B., Bzowski, M., Funsten, H. O., et al. 2021, ApJS, 254, 40

Schwadron, N. A., & McComas, D. J. 2013, ApJ, 764, 92

Sokół, J. M., McComas, D. J., Bzowski, M., & Tokumaru, M. 2020, ApJ, 897, 179

Swaczyna, P., Dayeh, M. A., & Zirnstein, E. J. 2023, ApJ, under review

Swaczyna, P., Bzowski, M., & Sokół, J. M. 2016, ApJ, 827, 71

Swaczyna, P., Eddy, T. J., Zirnstein, E. J., et al. 2022, ApJS, 258, 6

Swaczyna, P., McComas, D. J., Zirnstein, E. J., et al. 2020, ApJ, 903, 48

Schwadron et al. 2009, SSRv, 146, 207

Schwadron, N. A., Allegrini, F., Bzowski, M., et al. 2011, ApJS, 731, 56

Schwadron, N. A., and McComas, D. J. 2013, ApJ, 764, 92

Schwadron, N. A., Moebius, E., Fuselier, S. A., et al. 2014, ApJS, 215, 13

Schwadron, N. A., Allegrini, F., Bzowski, M., et al. 2018, ApJS, 239, 1

McComas, D. J., Barraclough, B. L., Funsten, H. O., et al. 2000, JGR, 105,10419

McComas, D. J., Elliott, H. A., Schwadron, N. A. et al. 2003, GRL, 30(10)

McComas, D. J., Allegrini, F., Bochsler, P., et al. 2009a, SSRv, 146, 11

McComas, D. J., Allegrini, F., Bochsler, P., et al. 2009b, Sci, 326, 959

McComas, D. J., Lewis, W. S., & Schwadron, N. A. 2014, Rev Geophys, 52, 118

McComas, D. J., Bzowski, M., Dayeh, M. A., et al. 2020, ApJS, 248, 26

McComas, D. J., Christian, E. R., Schwadron, N. A., et al. 2018, SSRv, 214, 116

McComas, D. J., Zirnstein, E. J., Bzowski, M., et al. 2017, ApJS, 229, 41

McComas, D. J., Dayeh, M. A., Funsten, H. O., et al. 2019, ApJ, 872, 127

Zirnstein, E. J., Heerikhuisen, J., McComas, D. J., et al. 2013, ApJ, 778, 112

Zirnstein, E. J., Heerikhuisen, J., Funsten, H. O., et al. 2016b, ApJL, 818, L18

Zirnstein, E. J., Heerikhuisen, J., & McComas, D. J. 2015, ApJL, 804, L22

Zirnstein, E. J., Heerikhuisen, J., McComas, D. J., & Schwadron, N. A. 2013, ApJ, 778, 112

Zirnstein E. J., T. K. Kim , M. A. Dayeh, et al. 2022a, ApJL, 937, L38

Zirnstein, E. J., Heerikhuisen, J., Zank, G. P., et al. 2017, ApJ, 836, 238

Zirnstein, E. J., Swaczyna, P., Dayeh, M. A., & Heerikhuisen, J. 2023, ApJ, in press

Zirnstein, E. J., Shrestha, B. L., McComas, D. J., et al. 2022b, NatAs, 6, 1398

Zirnstein, E. J., Giacalone, J., Kumar, R., et al. 2020, ApJ, 888, 29